\documentclass[preprint,12pt]{elsarticle}

\usepackage{amssymb}
\usepackage{wrapfig}

\usepackage{lineno}

\journal{Nuclear Instruments and Methods}

\begin{document}

\begin{frontmatter}



\title{Status of Air Shower Simulations}


\author{S.~Ostapchenko}

\address{NTNU, Institut  for fysikk, 7021 Trondheim, Norway}
\address{D.V. Skobeltsyn Institute of Nuclear Physics, Moscow
State University, 119992 Moscow, Russia}

\begin{abstract}
The present status of extensive air shower (EAS) simulation procedures is reviewed.
The advantages of combining numerical and Monte Carlo methods for the
description of EAS development are discussed.
Physics content of cosmic ray interaction models is briefly described
and their predictions are compared to the first LHC data.
 Finally, some outstanding puzzles related to  
cosmic ray composition at the ``ankle'' energies are analyzed. 

\end{abstract}
\end{frontmatter}


\section{Introduction}
\label{Sec:intro}
Over the last two decades, simulations of extensive air shower (EAS)
development have become an important ingredient of experimental analysis
of high energy cosmic ray (CR) data. The complexity of the corresponding
 procedures
is related to the fact that measured EAS characteristics have a very indirect
relation to the properties of the primary CR particles, resulting from a 
multi-step nuclear-electro-magnetic cascade development.
Air shower   simulations   can thus be improved in two directions:
i) towards higher precision and/or  efficiency of the calculations,
ii) regarding the physics content.

Concerning the former, applying EAS simulation procedures to showers induced
by ultra-high energy cosmic rays (UHECR),
 one faces the problem of enormous calculation
time required. Thus, one has to care about an optimization of the shower
 modeling,
in order to obtain sufficiently high simulation statistics, and about keeping
a high accuracy for calculations of both average EAS characteristics and 
 their distributions.

Coming to the latter, of special importance is a correct description of the
cascade of nuclear interactions of the hadronic component of air showers,
which acts as a backbone of EAS. On the other hand, the corresponding 
theoretical models
remain just phenomenological ones and involve relatively large number
of adjustable parameters. Thus, further model development related both to
improvements of the corresponding theoretical description and to retuning
of model parameters with new accelerator data is desirable.

\section{EAS simulation procedures}
\label{Sec:EAS}
The most transparent   approach to EAS simulation is the direct
modeling of the   shower development, as realized e.g.~in the CORSIKA
program \cite{hec98}: The propagation in the atmosphere and interactions
of each particle are traced using Monte Carlo (MC) methods.
 The approach has a natural
restriction: With the number of cascade particles rising proportionally
to the energy of the primary cosmic ray, so does the computing time
 required.
Hence, extension of the procedure to  very high energies requires
a certain optimization of the method, e.g.~employing weighted-sampling:
Only a number of representative particles among all the secondaries produced
per interaction is traced further by a code; each of those particles acquires
thus some weight.

 A classical example is the Hillas's ``thinning'' method
\cite{hil97}, where a single particle per interaction is chosen with the
probability $E_s/E_p$, $E_s$ being the energy of the given secondary and 
 $E_p$ -- the one of the primary particle. Correspondingly, the weight of the 
 chosen secondary is $w_s=E_p/E_s\; w_p$, $w_p$ being the weight of the primary.
 Though the method works well for average EAS characteristics, it introduces
 artificial fluctuations in the distributions of air shower observables,
 as discussed e.g.~in \cite{kob01}. The  solution to the problem
 was to impose a restriction on maximal  weights   \cite{kob01},
 in order to reduce the magnitude of  artificial fluctuations,
 and to complement the method by an ``unthinning'' procedure \cite{bil08}.
 The latter  allows one  to convert the distribution of ``weighted'' 
 particles
 coming from the ``thinned'' EAS simulation to  a realistic 
 particle distribution
  and to reduce  artificial fluctuations
 to a tolerable level. However, the simulation procedure
 remains   time-consuming -- as one  has to 
find a balance between a sparser ``thinning''
  and not too large weights.

On the other hand, the efficiency and the accuracy of air shower modeling
can be significantly improved combining MC and numerical methods, as is done
in the CONEX \cite{conex} and SENECA \cite{seneca} codes. 
Indeed, as EAS  fluctuations   are   dominated by interactions  and
the propagation in the atmosphere  of both the primary CR particle and of 
first few generations of the most energetic  secondaries, one can apply a
two-step procedure: Explicit MC simulation of the initial stage of the
shower and numerical description of secondary hadronic and electro-magnetic
(e/m) cascades, based on the solutions of the corresponding cascade
 equations.\footnote{In SENECA, the numerical solution is employed for 
 hadronic cascades while a pre-tabulation is used for e/m cascades.} 
For a number of applications, like calculations of fluorescence and
Cherenkov radiation profiles of air showers,
 the so-obtained one-dimensional EAS modeling is sufficient.
 In the more general case, when one is interested in the signal in 
 ground-based detectors, a three-step procedure is applied   \cite{seneca},
 which includes MC modeling of both the highest and the lowest energy part
 of EAS while intermediate energy range is described numerically.
 Treating   most of the cascade with numerical methods,
 one improves the efficiency of the procedure and enhances its accuracy.



\section{Hadronic interaction models}
\label{Sec:hadr}
As discussed in the Introduction, the least certain part of EAS simulation
procedures is the treatment of hadronic cascades in the atmosphere, 
which involves phenomenological models of hadronic and nuclear interactions.

Contemporary CR interaction models, like EPOS \cite{epos}, QGSJET \cite{qgs}
and QGSJET-II \cite{qgs2}, and SIBYLL \cite{sibyll}, are characterized by a
similar physics content, being designed to treat general hadronic collisions,
which involves both nonperturbative ``soft'' and ``hard'' parton processes.
Soft physics is usually described within the Reggeon Field Theory framework
 as soft Pomeron exchanges; hard parton dynamics is treated within
the DGLAP formalism and implemented in the models either following the so-called
minijet approach \cite{mini} or   the qualitatively similar
 ``semihard Pomeron'' scheme \cite{sh-pom}. Despite these general similarities,
 the models diverge in their predictions, which is both due to technical
 differences in the implementation of the above-discussed approaches and,
 especially, due to different treatments of non-linear interaction effects
 related to parton shadowing and saturation.

Around the energy of the CR knee, EAS characteristics obtained using different
models are relatively close to each other. This is both due to the model
calibration to  similar sets of accelerator data and due to ongoing model
tests by air shower experiments, notably by KASCADE \cite{ape07}, which
resulted in serious improvements of certain models. In the discussed energy
range, the observed EAS characteristics are rather well reproduced by
simulations under reasonable assumptions on the CR
composition, although certain contradictions  persist  and
 composition studies bear model-dependence \cite{ant05}.

The situation changes drastically at higher energies ($E_0>10^{18}$ eV),
where model predictions  diverge noticeably
and certain important observations by air shower experiments can not be
explained by the present models. 

\section{First measurements at the Large Hadron Collider}
\label{Sec:lhc}
The first LHC data have a strong impact on EAS simulation procedures and
on the interpretation of CR observations. Apart from providing
\begin{wrapfigure}{r}{.45\textwidth}
\includegraphics[width=6cm,height=5.5cm]{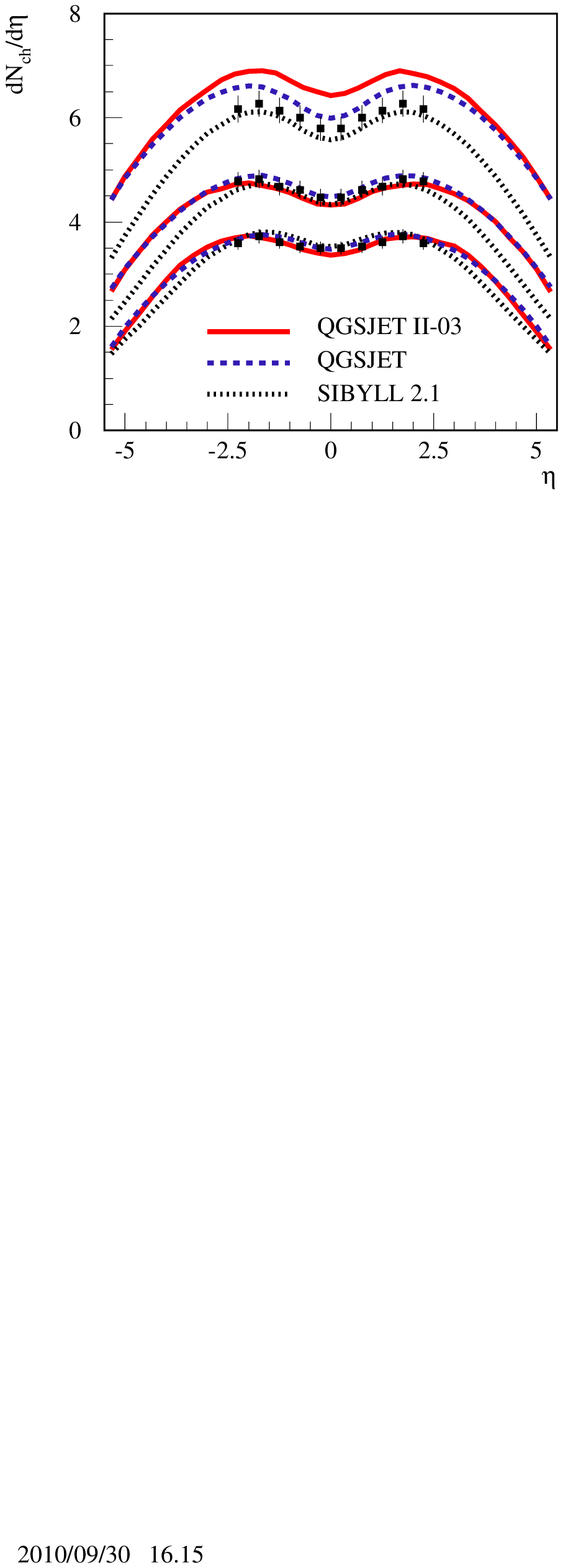}
\vspace{-0.1cm}
\caption{Predictions of CR interaction models for $dN_{\rm ch}/d\eta$
for different c.m.~energies (from up to bottom: $\sqrt s=7$, 2.38, 0.9 TeV)
compared to CMS data \cite{cms}.\label{fig:cms}}
\vspace{-0.5cm}
\end{wrapfigure}
 additional 
constraints for hadronic interaction models, measurements of secondary particle
production by the CMS and ALICE collaborations  gave no evidence
on a more rapid energy rise of the multiplicity of hadronic collisions 
than predicted  by the present CR interaction models.
This is illustrated in Fig.~\ref{fig:cms},
 where the CMS data on the pseudorapidity
density of charged particles in $pp$ collisions
 are compared to the corresponding results
of MC models, the latter being obtained applying the experimental trigger
to the MC generated hadronic final states. Clearly, the  data  are 
bracketed by the  model predictions. 

\section{UHECR puzzles}
\label{Sec:uhecr}
Historically, the first strong indication on a much higher EAS  muon content
than  predicted by simulations has been observed in the HiRes-Mia analysis
\cite{abb05} for CR energies $E_0>10^{17}$ eV. More recently, the Pierre Auger 
collaboration has demonstrated using 4 different methods
 that experimental data favor
 a much higher (by a factor $\sim 1.5$) number of muons $N_{\mu}$ at
ground than predicted e.g.~by the QGSJET-II model \cite{auger-nmu}.
Such a strong  $N_{\mu}$ enhancement can not be achieved with the
present CR interaction models, as it would require to increase the multiplicity
of proton-air and pion-air collisions by an order of magnitude {\it over 
a wide range of energies} \cite{prag,ulrich}. On the other hand, the data
are marginally consistent with simulation results for  iron-induced air 
showers \cite{auger-nmu}. As the measurements
refer  mainly to primary energies below 10 EeV, the required change to
 an iron-dominated
composition has to occur {\em around the ankle} of the CR spectrum. 

Another striking result obtained  was the sharp
decrease of the width of the shower maximum distribution  RMS($X_{\max}$) at 
$E_0>10^{18}$ eV \cite{auger-xmax}.
As discussed in  \cite{alo08}, RMS($X_{\max}$) is an almost model-independent
measure of the CR composition.
 Indeed, for proton-induced EAS this quantity is mainly
defined by the mean free pass of the proton 
$\lambda _p\sim 1/\sigma ^{\rm inel}_{p-{\rm air}}$, which sets the lower limit
on the  RMS($X_{\max}^p$) around 50 g/cm$^2$. Fluctuations related to the
geometry of $p$-air interactions (higher/smaller inelasticity for
 ``central''/peripheral collisions)
can only increase the corresponding value. On the other hand, 
in case of Fe-induced EAS, RMS($X_{\max}^{\rm Fe}$) is rather dominated 
by the fluctuations of the collision geometry,
primarily, via the variations of the number of ``wounded'' projectile nucleons
(which participate in particle production) and via the fragmentation of the
nuclear spectator part \cite{kal93}. Even for extreme assumptions,
  RMS($X_{\max}^{\rm Fe}$) can not exceed some 30  g/cm$^2$.  

  The observed decrease of RMS($X_{\max}$) from $\sim$55 g/cm$^2$ at 1 EeV
    to  $\sim$30  g/cm$^2$ at 30 EeV   may be naively interpreted as a 
   change from a $p$-dominated to an Fe-dominated composition in the
   discussed energy range  \cite{auger-xmax}. In reality, 
\begin{wrapfigure}{tr}{.45\textwidth}
\includegraphics[width=6cm,height=5.3cm]{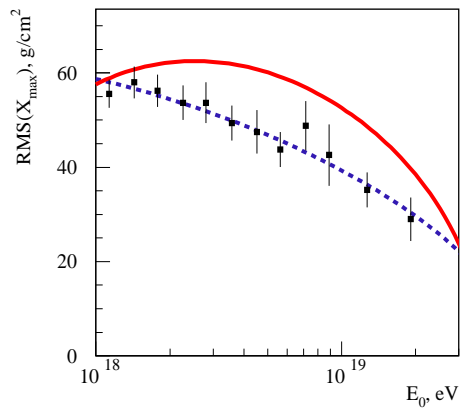}
\vspace{-0.1cm}
\caption{RMS($X_{\max}$) for an energy-dependent CR composition
(as described in the text): $f_p(1\:{\rm EeV})=1$ (solid) and
 $f_p(1\;{\rm EeV})=0.4$ (dashed); points -- Pierre Auger data
   \cite{auger-xmax}.\label{fig:rms}}
\vspace{-0.4cm}
\end{wrapfigure} 
as correctly noticed 
    in \cite{taylor,wilk}, adding a small admixture of iron nuclei  to the pure proton
   composition increases the width of the  distribution.
   As an illustration,  in Fig.~\ref{fig:rms}
    we compare the  Pierre Auger data on  
    RMS($X_{\max}$) with EAS simulation results (using QGSJET-II) considering
    a simple 2-component ($p+Fe$)
     composition and assuming the partial abundances $f_i$ to 
     change smoothly between 1 and 30 EeV:
    $ f_p(E)=f_p(1\:{\rm EeV})\left[1- \lg (E/{\rm EeV})/1.5\right]$,
    $f_{\rm Fe}(E)=1-f_p(E)$.
    Comparing the cases $f_p(1\;{\rm EeV})=1$ and $f_p(1\:{\rm EeV})=0.4$,
    we see that it is the latter choice which is supported by the data,
    i.e.~iron dominance at the ankle is favored.

     On the other hand, a heavy CR composition in the  EeV energy range
    is at variance  with HiRes and Pierre Auger data on the
     average  $X_{\max}$, which are consistent with
    the proton-dominance. In order to reconcile the latter
     with a heavy composition, a much deeper (than presently
    predicted) shower    penetration in the atmosphere has to be assumed.
    In practical terms this would require a factor of 2 decrease for 
    $\sigma ^{\rm inel}_{p-{\rm air}}$ compared to the current model 
    predictions \cite{ulrich}, which
     would correspond to a similar reduction for 
    the total $pp$ cross section
      at $\sqrt s\sim 6$ TeV, i.e.~only slightly
    above the Tevatron energy.  Such a sudden fall down of 
    $\sigma ^{\rm tot}_{pp}$ would imply  very exotic physics.

\section{Conclusions}
\label{Sec:conc}
Significant progress in the  modeling of air showers
has been achieved in recent years. Combing  numerical and MC methods,
 as realized in the CONEX and SENECA codes,
allowed one to increase  the efficiency and the accuracy
of EAS simulations.
The description of hadronic collisions by the corresponding MC
models has been considerably improved, particularly, concerning the treatment
of non-linear interaction effects. Predictions of various hadronic MC
models have   converged to each other, both due to the improved
theoretical description and due to model tests by EAS experiments.

The first LHC data on secondary hadron production in $pp$ collisions 
provide no indication on a more rapid energy rise of the multiplicity  
 than predicted by the present CR interaction models.
Thus,  a strong rise of   EAS muon content  at $1\div 100$ PeV is
 not supported by the collider observations.
 
The strong enhancement of EAS muon content and the sharp decrease of the 
width of   $X_{\max}$ distributions, as observed by the  Pierre Auger
collaboration above 1 EeV, can not be explained in the framework of the 
present CR interaction models, unless an iron dominance of the 
CR composition at the ankle  is assumed. The latter assumption is however in
a strong conflict with $X_{\max}$ measurements at   $1\div 10$ EeV by all
the experiments working in this energy range.
In turn, the data on average  $X_{\max}$ can not be  reconciled with a 
heavy CR composition without invoking very exotic physics.

\vspace{2mm}
\noindent
{\bf Acknowledgements}
The author acknowledges discussions with Michael Kachelriess and
a fellowship from the program Romforskning of Norsk
Forsknigsradet.

\end{document}